\documentclass{article}
\usepackage{spconf,amsmath,graphicx}
\usepackage{setspace}
\usepackage{newtxmath}
\usepackage{multirow}
\usepackage{cite}
\usepackage{hyperref}
\makeatletter
\newcommand{\thickhline}{%
	\noalign {\ifnum 0=`}\fi \hrule height 1.0pt
	\futurelet \reserved@a \@xhline
}

\ninept
\title{VCVTS: Multi-speaker Video-to-Speech synthesis via cross-modal knowledge transfer from voice conversion}
\name{Disong Wang$^{1*}$\thanks{*Work done during internship at Tencent AI lab}, Shan Yang$^2$, Dan Su$^2$, Xunying Liu$^1$, Dong Yu$^2$, Helen Meng$^{1,3}$}
\address{$^1$The Chinese University of Hong Kong, Hong Kong SAR, China\\$^2$Tencent AI lab, China\\$^{3}$Centre for Perceptual and Interactive Intelligence, Hong Kong SAR, China\\\small{\{dswang, xyliu, hmmeng\}@se.cuhk.edu.hk}, {\{shaanyang, dansu, dyu\}@tencent.com}}
\begin{document}
%\ninept
%
\maketitle
\begin{abstract}
Though significant progress has been made for speaker-dependent Video-to-Speech (VTS) synthesis, little attention is devoted to multi-speaker VTS that can map silent video to speech, while allowing flexible control of speaker identity, all in a single system. This paper proposes a novel multi-speaker VTS system based on cross-modal knowledge transfer from voice conversion (VC), where vector quantization with contrastive predictive coding (VQCPC) is used for the content encoder of VC to derive discrete phoneme-like acoustic units, which are transferred to a Lip-to-Index (Lip2Ind) network to infer the index sequence of acoustic units. The Lip2Ind network can then substitute the content encoder of VC to form a multi-speaker VTS system to convert silent video to acoustic units for reconstructing accurate spoken content. The VTS system also inherits the advantages of VC by using a speaker encoder to produce speaker representations to effectively control the speaker identity of generated speech. Extensive evaluations verify the effectiveness of proposed approach, which can be applied in both constrained vocabulary and open vocabulary conditions, achieving state-of-the-art performance in generating high-quality speech with high naturalness, intelligibility and speaker similarity. Our demo page is released here\footnote{Demo: \url{https://wendison.github.io/VCVTS-demo/}}.
\end{abstract}
%
% \vspace{-0.5em}
\begin{keywords}
Multi-speaker, Video-to-Speech synthesis, voice conversion, knowledge transfer, vector quantization
\end{keywords}
\vspace{-1em}
\section{Introduction}
\label{sec:intro}
\vspace{-0.5em}

Video-to-Speech (VTS) synthesis aims to reconstruct speech signals from silent video by exploiting their bi-modal correspondences. VTS has various compelling applications such as assistive communication for patients with inability to produce voiced sounds (i.e., aphonia), and voice restoration for videoconferencing when speech signals are corrupted by noise or lost. Previous efforts generally build statistical models to map visual features to acoustic features, then use a vocoder to synthesize the waveform. Typical modelling approaches include Hidden Markov Models \cite{milner2015reconstructing,hueber2016statistical}, non-negative matrix factorization \cite{aihara2015lip}, maximum likelihood estimation \cite{ra2017visual} and deep learning methods \cite{milner2015reconstructing,le2017generating,ephrat2017vid2speech,ephrat2017improved, takashima2019exemplar, kumar2018harnessing,akbari2018lip2audspec,michelsanti2020vocoder,kumar2019lipper,vougioukas2019video,mira2021end,yadav2021speech,prajwal2020learning,oneata2021speaker}. Most works \cite{milner2015reconstructing,hueber2016statistical,aihara2015lip,ra2017visual,le2017generating,ephrat2017vid2speech,ephrat2017improved, takashima2019exemplar,kumar2018harnessing,akbari2018lip2audspec,michelsanti2020vocoder,kumar2019lipper,vougioukas2019video,mira2021end,yadav2021speech} are restricted to small datasets (e.g., GRID \cite{cooke2006audio}) to create single-speaker systems under constrained conditions with limited vocabulary, which hinders their practical deployment. A few studies \cite{prajwal2020learning,oneata2021speaker} propose to use speaker representations to capture speaker characteristics and control the speaker identity of generated speech, such that multi-speaker VTS can be achieved in a single system. \cite{prajwal2020learning} uses a pre-trained speaker encoder optimized for speaker recognition task to extract speaker representations, \cite{oneata2021speaker} applies adversarial training to achieve speaker disentanglement. Both approaches directly map cropped lips to speech, which treats the deep neural model as a black box, leading to insufficient interpretability of intermediate representations learned by the model. Besides, compared with other speech generation tasks, e.g., text-to-speech (TTS) synthesis or voice conversion (VC), current VTS systems tend to generate less natural-sounding speech. 

This paper proposes a novel multi-speaker VTS system referred to as VCVTS (Voice Conversion-based Video-To-Speech), which provides a more legible mapping from lips to speech. This is accomplished by first converting lips to intermediate phoneme-like acoustic units, which are used to accurately restore the spoken content. Besides, VCVTS can generate high-quality speech with flexible control of the speaker identity, leveraging the capabilities of a VC system. Specifically, the proposed approach contains three components: (1) Training a VC system which contains four modules - a \textit{content encoder} using vector quantization with contrastive predictive coding (VQCPC) \cite{baevski2019vq} to derive discrete phoneme-like acoustic units, a \textit{speaker encoder} extracting effective speaker representations to control the speaker identity, a \textit{pitch predictor} inferring the fundamental frequency ($F_0$) to control the pitch contour, and a \textit{decoder} mapping acoustic units, speaker representation and $F_0$ to mel-spectrograms; (2) The discrete phoneme-like acoustic units are treated as knowledge, which is transferred across modalities from speech to image, by training a Lip-to-Index (Lip2Ind) network to predict the index sequence of acoustic units, where the index corresponds to the acoustic unit obtained from the codebook of vector quantization (VQ); and (3) A multi-speaker VTS system, i.e., VCVTS, is formed by concatenating the Lip2Ind network with the VQ-codebook, speaker encoder, pitch predictor and decoder of VC. VCVTS provides a legible mapping from lips to speech and inherits the advantages of VC to effectively control speaker identity and pitch contour, which leads to the generation of high-quality speech outputs. 

The main contributions of this work are: (1) Derivation of discrete phoneme-like acoustic units by using VQCPC under the VC framework; (2) Development of a Lip2Ind network via cross-modal knowledge transfer to map lips to acoustic units for reconstructing spoken content; and (3) Development of a novel multi-speaker VTS system that can process diverse, unconstrained vocabularies and complicated image scenarios, e.g., on LRW dataset \cite{chung2016lip}.

\begin{figure*}[t]
  \centering
  \centerline{\includegraphics[width=1.0\textwidth]{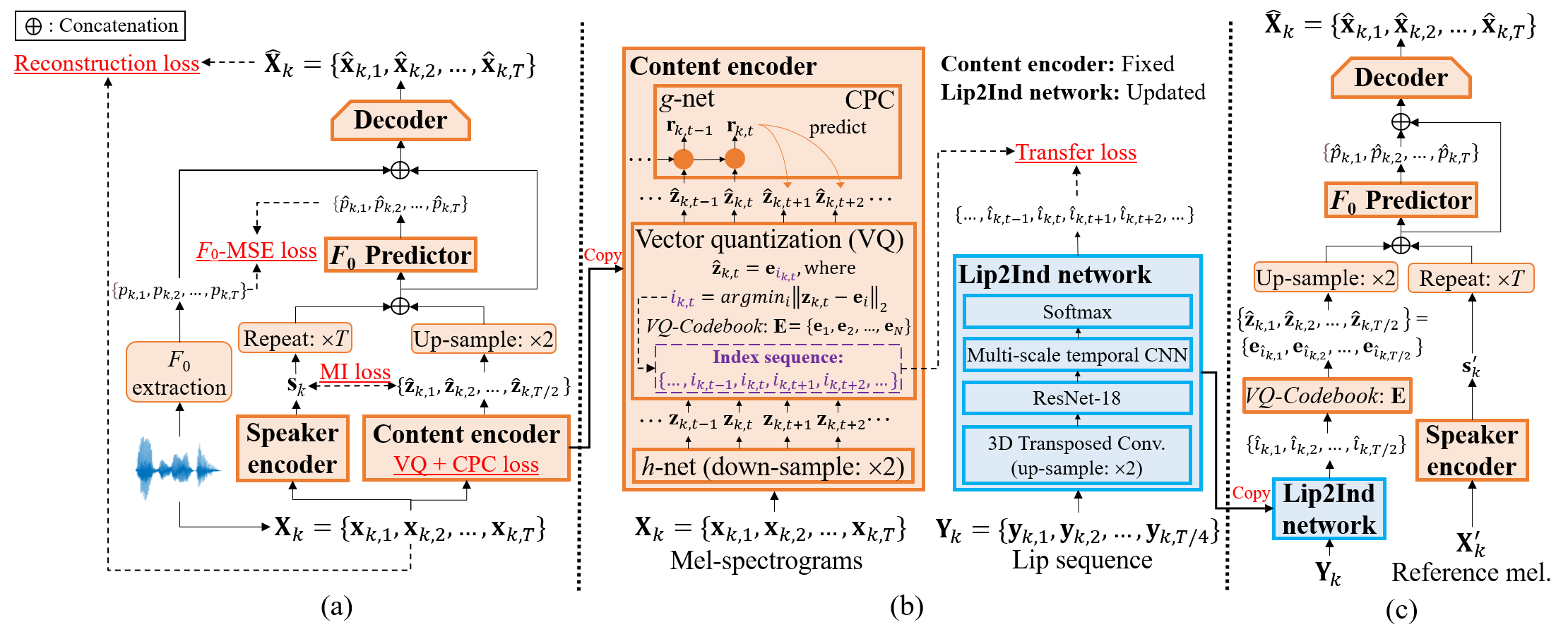}}
  \vspace{-2em}
  \caption{Diagram of the proposed approach: (a) VC system training by using VQCPC for the content encoder to derive discrete acoustic units; (b) Cross-modal knowledge transfer from the content encoder of VC to a Lip2Ind network; (c) A multi-speaker VTS system, i.e., VCVTS, which is formed by concatenating the Lip2Ind network with the VQ-codebook, speaker encoder, $F_0$ predictor and decoder of VC system.}\label{dia}
  \vspace{-1.5em}
\end{figure*}

\vspace{-1em}
\section{Proposed approach}
\label{sec:prop}
\vspace{-0.5em}
Fig. \ref{dia} shows the diagram of proposed approach, which contains three components that are elaborated in the following subsections. Assuming that there are \textit{K} videos with 25 frames per second (FPS), each video contains two streams: speech and image. For the $\textit{k}^{th}$ video, the speech stream is processed to mel-spectrograms $\textbf{X}_k=\{\textbf{x}_{k,1},\textbf{x}_{k,2},...,\textbf{x}_{k,T}\}$ computed with a shift size of 10ms (i.e., 100 FPS) to have a length of \textit{T}, the image stream is processed to obtain the lip sequence  $\textbf{Y}_k=\{\textbf{y}_{k,1},\textbf{y}_{k,2},...,\textbf{y}_{k,T/4}\}$ with a length of \textit{T}/4.   

\vspace{-0.8em}
\subsection{VC system training}
\vspace{-0.5em}
As shown in Fig. \ref{dia} (a), we propose a new VC system that is modified from VQMIVC \cite{wang21n_interspeech}. Four modules are contained: speaker encoder, content encoder, pitch predictor and decoder, where the first three modules produce the speaker representation, discrete acoustic units and $F_0$ respectively, which are fed into the decoder to reconstruct mel-spectrograms. The pitch predictor is newly added to infer $F_0$ values that are difficult to be predicted from visual features \cite{milner2015reconstructing}. VQCPC is used for content encoding as shown in Fig. \ref{dia} (b). The \textit{h}-net of content encoder first maps $\textbf{X}_k$ to $\textbf{Z}_k=\{\textbf{z}_{k,1},\textbf{z}_{k,2},...,\textbf{z}_{k,T/2}\}$ with a down-sampling factor of 2, then applies VQ on $\textbf{Z}_k$ to obtain the discrete acoustic units $\hat{\textbf{Z}}_k$=\{$\hat{\textbf{z}}_{k,1}$,$\hat{\textbf{z}}_{k,2}$,...,$\hat{\textbf{z}}_{k,T/2}$\}:
% \begin{footnotesize}
\begin{equation}
% \vspace{-0.75em}
    {\hat{\textbf{z}}_{k,t}}={{\textbf{e}}_{{{i}_{k,t}}}}, {{i}_{k,t}}=\arg {{\min }_{i}}{{\left\| {{\textbf{z}}_{k,t}}-{{\textbf{e}}_{i}} \right\|}_{2}} \label{ind}
% \vspace{-0.25em}
\end{equation}
% \end{footnotesize}
where $\textbf{e}_i$ is $i^{th}$ unit of the VQ-codebook $\textbf{E}=\{\textbf{e}_{1},\textbf{e}_{2},...,\textbf{e}_{N}\}$. To learn a content-related codebook $\textbf{E}$, we minimize the VQ loss \cite{van2020vector} to impose an information bottleneck:
% \begin{footnotesize}
\begin{equation}
% \vspace{-0.75em}
    {{L}_{VQ}}=\frac{2}{KT}\sum\limits_{k=1}^{K}{\sum\limits_{t=1}^{T/2} \left\| {{\textbf{z}}_{k,t}}-sg({{{\hat{\textbf{z}}}}_{k,t}}) \right\|_{2}^{2}} \label{vq-loss}
% \vspace{-0.25em}
\end{equation}
% \end{footnotesize}
where \textit{sg}(·) denotes the stop-gradient operator. Besides, a \textit{g}-net is added after $\hat{\textbf{Z}}_k$ to obtain $\textbf{R}_k=\{\textbf{r}_{k,1},\textbf{r}_{k,2},...,\textbf{r}_{k,T/2}\}$ to distinguish a positive sample $\hat{\textbf{z}}_{k,t+m}$ that is \textit{m} steps in the future from negative samples drawn from the set ${\Omega}_{k,t,m}$ by minimizing the contrastive predictive coding (CPC) loss \cite{wang21n_interspeech,oord2018representation}:
% \begin{footnotesize}
\begin{equation}
% \vspace{-0.75em}
    \!\!{{L}_{CPC}}=-\frac{1}{KT'M}\!\sum\limits_{k=1}^{K}{\sum\limits_{t=1}^{T'}{\sum\limits_{m=1}^{M}{\!\log\! \left[\! \frac{\exp (\hat{\textbf{z}}_{k,t+m}^{T}{{\mathbf{W}}_{m}}{{\textbf{r}}_{k,t}})}{\sum\nolimits_{\tilde{\textbf{z}}\in {{\Omega }_{k,t,m}}}{\exp ({{{\tilde{\textbf{z}}}}^{T}}{{\mathbf{W}}_{m}}{{\textbf{r}}_{k,t}})}} \!\right]}}} \label{cpc-loss} 
% \vspace{-0.25em}
\end{equation}
% \end{footnotesize}
where ${T'}=T/2-M$, $\textbf{W}_m$ (\textit{m}=1,2,...,\textit{M}) is a trainable projection matrix. CPC forces the discrete acoustic units $\hat{\textbf{Z}}_k$ to capture `slow features' \cite{wiskott2002slow} that span many time steps, e.g., phonemes. To separate $\hat{\textbf{Z}}_k$ from speaker representation $\textbf{s}_k$, their mutual information (MI) ${L}_{MI}$ \cite{gierlichs2008mutual,cheng2020club} is estimated and minimized during the training. Together with the $F_0$ prediction loss ${{L}_{{{F}_{0}}-MSE}}$ and mel-spectrograms reconstruction loss ${{L}_{REC}}$, the total loss used for training the VC system is given in Eq. (\ref{vc-loss}). More training details can be found in \cite{wang21n_interspeech}.
\begin{equation}
% \vspace{-2em}
    {{L}_{VC}}={{L}_{VQ}}+{{L}_{CPC}}+{{L}_{MI}}+{{L}_{{{F}_{0}}-MSE}}+{{L}_{REC}} \label{vc-loss}
% \vspace{-0.25em}
\end{equation}

\begin{figure}[t]
  \centering
  \centerline{\includegraphics[width=0.5\textwidth]{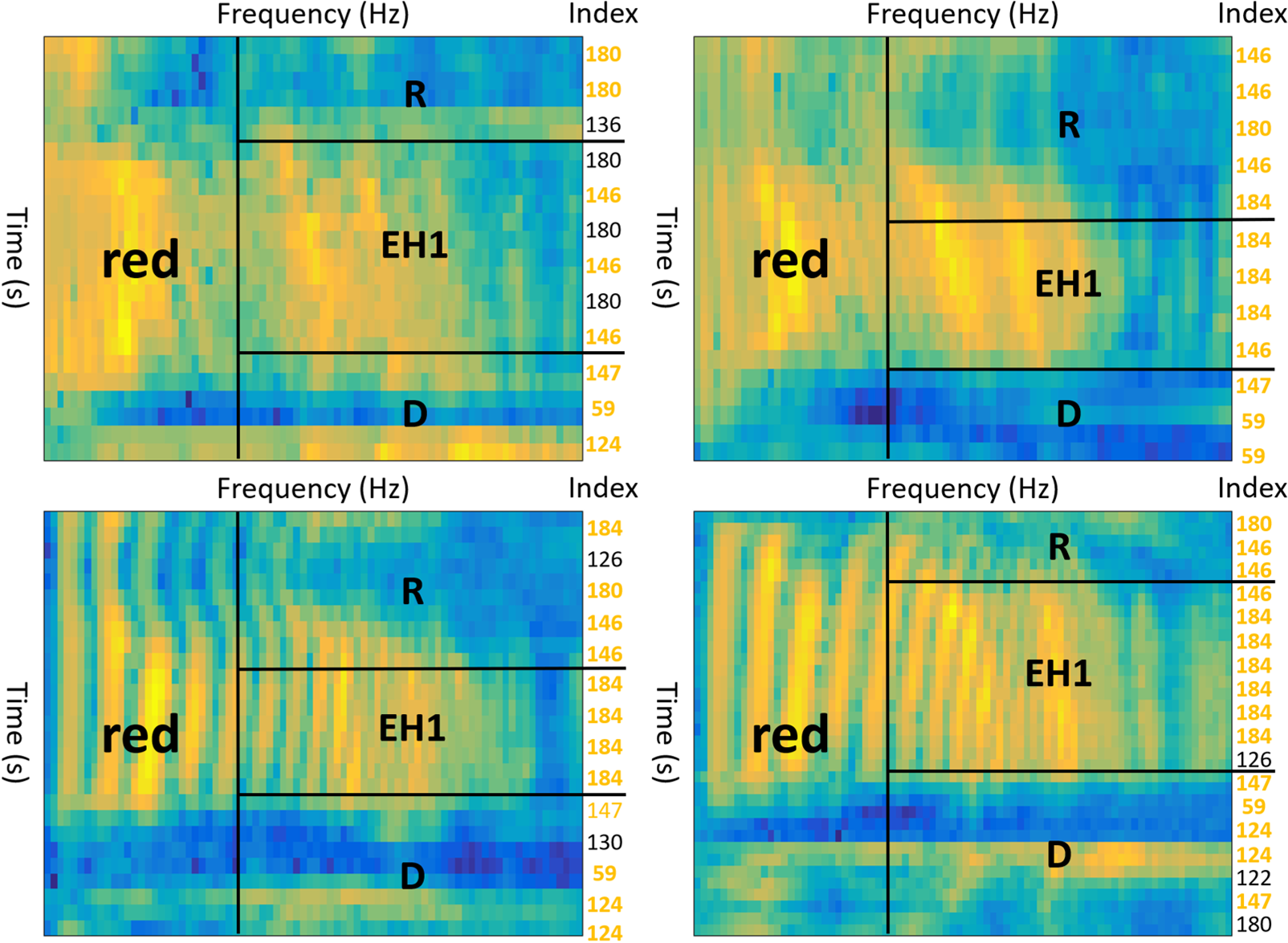}}
  \vspace{-1em}
  \caption{Visualization of mel-spectrograms and acoustic units (denoted by the index sequence) derived by the content encoder of VC from four speakers (selected from the GRID dataset) uttering the same word `red' with phoneme transcription `R EH1 D', the common index that occurs at least twice in the four index sequences within the same phoneme is marked as orange.}\label{au}
  \vspace{-1.5em}
\end{figure}

\begin{table*}[t]
  \caption{Objective and subjective evaluation results of different VTS systems on testing speakers, where `Seen' and `Unseen' denote that testing speakers are respectively seen and unseen during training, and subjective results are MOS with 95\% confidence intervals for Speech Naturalness (MOS-SN) and Speaker Similarity (MOS-SS). }
  \label{os}
%   \vspace{-1em}
  \centering
  \scalebox{0.98}{
  \begin{tabular}{c|c|c|c|c|c|c|c|c|c|c}
    \thickhline
    \multirow{2}{*}{Dataset} & \multirow{2}{*}{System} &  \multirow{2}{*}{Vocoder} & \multirow{2}{*}{Speakers} & \multicolumn{5}{c}{Objective} & \multicolumn{2}{|c}{Subjective} \\
    \cline{5-11}
    & & & & PESQ $\uparrow$ & STOI $\uparrow$ & ESTOI $\uparrow$ & MCD $\downarrow$ & $F_0$-RMSE $\downarrow$ & MOS-SN $\uparrow$ & MOS-SS $\uparrow$  \\
    \hline
    \multirow{8}{*}{GRID} & XTS \cite{oneata2021speaker} & GL & Seen & 1.471 & 0.494 & 0.268 & 8.52 & 55.24 & 2.59 $\pm$ 0.09 & 3.56 $\pm$ 0.11 \\
    % \cline{2-11}
    & Lip2Wav \cite{prajwal2020learning} & GL & Seen & 1.690 & 0.610 & 0.434 & 7.59 & 49.33 & 3.41 $\pm$ 0.11 & 4.30 $\pm$ 0.11  \\
    % \cline{2-11}
    & VCVTS (ours) & GL & Seen & \textbf{1.816} & \textbf{0.691} & \textbf{0.512} & 6.15 & \textbf{43.16} & 3.46 $\pm$ 0.12 & 4.38 $\pm$ 0.11 \\
    % \cline{2-11}
    & VCVTS (ours) & PWG & Seen & 1.670 & 0.673 & 0.481 & \textbf{5.23} & 52.14 & \textbf{3.78 $\pm$ 0.10} & \textbf{4.47 $\pm$ 0.12} \\
    \cline{2-11}
    & XTS \cite{oneata2021speaker} & GL & Unseen & 1.295 & 0.439 & 0.160 & 12.04 & 88.97 & 2.33 $\pm$ 0.11 & 1.99 $\pm$ 0.18 \\
    % \cline{2-11}
    & Lip2Wav \cite{prajwal2020learning} & GL & Unseen & 1.293 & 0.419 & 0.202 & 10.11 & 55.35 & 3.13 $\pm$ 0.12 & 2.44 $\pm$ 0.16  \\
    % \cline{2-11}
    & VCVTS (ours) & GL & Unseen & \textbf{1.417} & \textbf{0.582} & \textbf{0.330} & 8.36 & \textbf{50.80} & 3.25 $\pm$ 0.11 & 2.66 $\pm$ 0.17 \\
    % \cline{2-11}
    & VCVTS (ours) & PWG & Unseen & 1.386 & 0.581 & 0.326 & \textbf{7.30} & 54.60 & \textbf{3.60 $\pm$ 0.10} & \textbf{2.75 $\pm$ 0.16} \\
    \hline
    \multirow{3}{*}{LRW} & Lip2Wav \cite{prajwal2020learning} & GL & Unseen & 1.197 & 0.543 & 0.344 & 9.18 & 71.29 & 2.65 $\pm$ 0.10 & 3.05 $\pm$ 0.18  \\
    % \cline{2-11}
     & VCVTS (ours) & GL & Unseen & \textbf{1.352} & \textbf{0.628} & \textbf{0.458} & 7.34 & \textbf{47.76} & 3.22 $\pm$ 0.11 & 3.68 $\pm$ 0.13 \\
    % \cline{2-11}
    & VCVTS (ours) & PWG & Unseen & 1.281 & 0.608 & 0.429 & \textbf{6.81} & 55.72 & \textbf{3.38 $\pm$ 0.12} & \textbf{3.72 $\pm$ 0.14} \\
    \thickhline
  \end{tabular}
  }
  \vspace{-1.5em}
\end{table*}

% \vspace{-1.0em}
\vspace{-0.8em}
\subsection{Cross-modal knowledge transfer}
\vspace{-0.5em}
We illustrate some acoustic units derived by a well-trained content encoder of VC as shown in Fig. \ref{au}, the acoustic units are denoted by the index sequence $\{i_{k,t}\}$ that is obtained by Eq. (\ref{ind}). We can observe that same indices tend to appear within the same phoneme, and different indices tend to occur in different phonemes, this indicates that the index sequence is strongly associated with the underlying linguistic content, e.g., phonemes. Therefore, the knowledge, i.e., index, can be transferred to the VTS task, such that the accurate spoken content can be appropriately inferred from lip features.

Therefore, for the VQ-codebook \textbf{E} containing \textit{N} units, a Lip2Ind network is designed to infer indices from lip sequence as shown in Fig. \ref{dia} (b), which is a \textit{N}-way classification problem. Given the video with mel-spectrograms $\textbf{X}_k$ and lip sequence $\textbf{Y}_k$, the content encoder takes in $\textbf{X}_k$ to obtain the index sequence $\{{i}_{k,1},{i}_{k,2},...,{i}_{k,T/2}\}$, which is treated as the learning target of the Lip2Ind network that takes $\textbf{Y}_k$ as the input. So the Lip2Ind network is trained to minimize the following knowledge transfer loss (i.e., cross-entropy):
% \begin{footnotesize}
\begin{equation}
% \vspace{-0.75em}
    {{L}_{Trans}}=-\frac{2}{KT}\sum\limits_{k=1}^{K}{\sum\limits_{t=1}^{T/2}{\log  q_{_{{{i}_{k,t}}}} }}
% \vspace{-0.25em}
\end{equation}
% \end{footnotesize}
where $q_{_{{{i}_{k,t}}}}$ is the output of the Lip2Ind network at time \textit{t} and denotes the posterior probability of index ${i}_{k,t}$.

\vspace{-0.5em}
\subsection{Multi-speaker VTS system}
\vspace{-0.5em}
The proposed VCVTS is shown in Fig. \ref{dia} (c), where the well-trained Lip2Ind network with VQ-codebook can produce accurate acoustic units for spoken content reconstruction, so it can be concatenated with the speaker encoder, pitch predictor and decoder of VC to form a multi-speaker VTS system. This configuration leverages the power of VC to generate the speech with the speaker identity controlled by effective speaker representation $\textbf{s}'_k$, which is produced by the speaker encoder from a reference utterance, and pitch contour that is controlled by realistic $F_0$ values inferred by the pitch predictor.

\vspace{-1em}
\section{Experiments}
\label{sec:typestyle}
\vspace{-0.5em}
Experiments are separately conducted on GRID \cite{cooke2006audio} and LRW \cite{chung2016lip} datasets. GRID contains 33 speakers with 1K videos per speaker, the vocabulary has 52 words which are recorded in a constrained indoor condition. We use 4 speakers (s1, s2, s4 and s29) as testing speakers, and consider two partitions of speakers used for training: all 33 speakers and 29 speakers excluding 4 testing speakers, which measures the performance of VTS for in-domain and out-of-domain speakers, respectively. Videos of each speaker are randomly split into train/dev/test with ratios of 80\%/10\%/10\%. LRW is obtained in unconstrained outdoor conditions, with the vocabulary more than 500 words spoken by hundreds of speakers. We use the default splits of train/dev/test of LRW for experiments, and assume that speakers in different splits are exclusive considering that the speaker label of each video is unavailable. For speech features, 80-dim mel-spectrograms are calculated with 400-point fast Fourier transform and 10ms shift size that is also used for extracting $\textit{F}_0$. For image features, Dlib toolkit \cite{dlib09} is employed to obtain facial landmarks that are used to crop and resize the lip region to have a size of 96$\times$96.  

The VC system follows most settings of VQMIVC \cite{wang21n_interspeech}, except that the speaker encoder is taken from Resemblyzer\footnote{\url{https://github.com/resemble-ai/Resemblyzer}} and outputs a 256-dim vector, the VQ-codebook of content encoder has 200 160-dim units, $F_0$ predictor contains 2-layer 1D-convolutional network with channel of 384, kernel size of 3, ReLU and layer normalization, which is followed by a linear layer to predict 1-dim $F_0$ values. Besides, the decoder of VC is based on Conformer \cite{gulati2020conformer}, which consists of 4 blocks, each block has 384-dim attention, 2 attention heads, 1536-dim feed forward (FFN) layers, and 31x1 convolutional kernel size. The speech streams of GRID and LRW are separately used to train different VC systems by Adam \cite{kingma2014adam} for 150 epochs with the first 15-epoch warmup increasing the learning rate from 1e-6 to 1e-3, and batch size of 256. The Lip2Ind network is modified from \cite{martinez2020lipreading}, and consists of a 3D-transposed convolutional network to up-sample lip sequence with a factor of 2, a ResNet-18 \cite{he2016deep}, a 4-layer multi-scale temporal CNN with each temporal convolution composed of 3 branches that have the kernel size of 3, 5 and 7 respectively, and a softmax layer to predict the probability distribution of index. The Lip2Ind network is trained by Adam for 80 epochs using a cosine scheduler with the initial learning rate of 3e-4, a weight decay of 1e-4 and batch size of 32, where we also use data augmentation including random cropping of 88$\times$88 pixels, random horizontal flip and mixup \cite{zhang2017mixup}. For the proposed VCVTS, mel-spectrograms are converted to waveform by using Griffin-Lim (GL) \cite{griffin1984signal} or Parallel WaveGAN (PWG) \cite{yamamoto2020parallel} vocoder that is trained on LibriTTS \cite{zen2019libritts}. We compare VCVTS with two multi-speaker VTS baseline systems, i.e., Lip2Wav \cite{prajwal2020learning} and XTS \cite{oneata2021speaker}, both of which use GL following their papers, where XTS is only implemented on GRID as its training requires speaker labels that are unavailable in LRW.

\begin{figure*}[t]
  \centering
  \centerline{\includegraphics[width=0.88\textwidth]{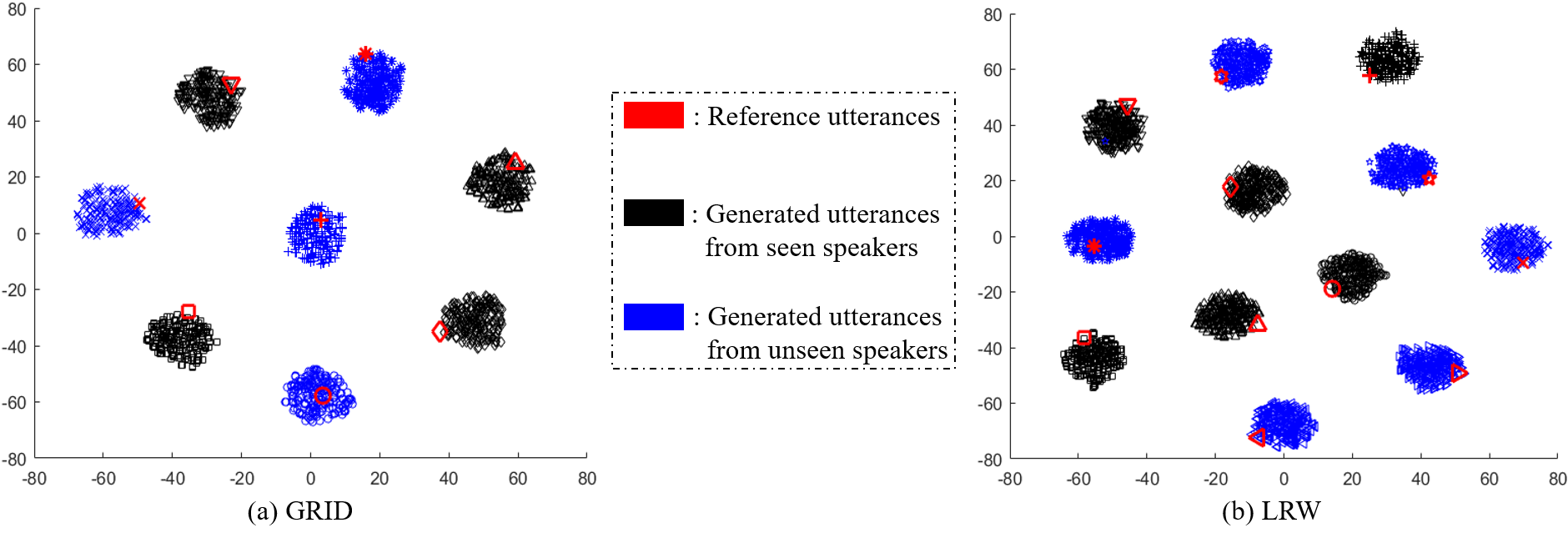}}
  \vspace{-1.5em}
  \caption{Visualization of speaker representations of reference and generated utterances, where reference utterances from different speakers are highlighted with larger, red markers with different shapes, generated utterances are highlighted with smaller markers and the same shape as their corresponding reference utterance that is used to control the speaker identity.}\label{spk-vis}
  \vspace{-1.5em}
\end{figure*}

\vspace{-0.5em}
\subsection{Objective evaluation results}
\vspace{-0.5em}
Following previous works \cite{prajwal2020learning,oneata2021speaker}, we adopt three standard speech quality metrics: perceptual evaluation of speech quality (PESQ) \cite{rix2001perceptual} to measure the quality, Short-Time Objective Intelligibility (STOI) \cite{taal2010short} and Extended STOI (ESTOI) \cite{jensen2016algorithm} to measure the speech intelligibility. We also compute mel-cepstral Distortion (MCD) and Root Mean Square Error of $F_0$ ($F_0$-RMSE). Table \ref{os} shows the objective evaluation results of different VTS systems on testing speakers that are seen or unseen during training. 

For the GRID, we can observe that compared with results for the unseen testing speakers, all VTS systems achieve better performance on the seen testing speakers by a great margin, this shows the susceptibility of VTS systems to out-of-domain speakers. The proposed VCVTS using GL achieves highest scores for both seen and unseen speakers on PESQ, STOI, ESOI and $F_0$-RMSE, which shows that the Lip2Ind network can accurately infer the indices of discrete acoustic units for generating the speech with higher quality and intelligibility, and the proposed pitch predictor can be used to infer realistic $F_0$ values to control the pitch contour to be closer to that of original speech. Besides, although the proposed VCVTS using the PWG vocoder outperforms baseline systems on most metrics, it is inferior to the proposed VCVTS using GL except for MCD, as PWG is based on neural networks and tends to introduce acoustic artifacts that degrade the objective performance. For the LRW, it is encouraging to see that VCVTS using GL or PWG outperforms the start-of-the-art VTS system, i.e., Lip2Wav, this indicates that under unconstrained conditions, the proposed system can still employ the Lip2Ind network and VQ-codebook to obtain accurate acoustic units used to restore the spoken content. Consequently, the generated speech has more accurate spectral details and pitch contour.  

\vspace{-0.5em}
\subsection{Subjective evaluation results}
\label{ser}
\vspace{-0.5em}
Subjective listening tests have been conducted to give mean opinion score (MOS) for Speech Naturalness (MOS-SN) and Speaker Similarity (MOS-SS). 20 subjects are recruited to give 5-point MOS, i.e.,  1-bad, 2-poor, 3-fair, 4-good, 5-excellent. For both GRID and LRW, 60 generated utterances are randomly selected from different systems for MOS tests, where 15 utterances are selected for each of 4 testing speakers of GRID, and the results are reported in Table \ref{os}. 

For the GRID, we observe that compared with results on seen testing speakers, the performance on unseen testing speakers of all VTS systems deteriorates with significant degradation of speaker similarity, which is caused by the limited number of speakers of GRID for training a VTS system that generalizes poorly to out-of-domain speakers. This problem can be alleviated for LRW with hundreds of speakers used for training, where both Lip2Wav and VCVTS achieve higher MOS of speaker similarity on LRW than that on GRID. Besides, We can see that VCVTS still outperforms the Lip2Wav and XTS with higher MOS-SN and MOS-SS, this shows that the proposed VCVTS inherits the advantages of VC to generate the speech with higher naturalness and voice similarity. Besides, for the proposed VCVTS, PWG is superior to GL, this shows that PWG tends to generate the waveform that sounds more realistic, which improves the perceptual quality of human.

\vspace{-0.5em}
\subsection{Visualization of speaker representation}
\vspace{-0.5em}
To further contextualize the speaker similarity evaluation results illustrated in section \ref{ser}, we select the reference utterances from different speakers that are seen or unseen during training to control the speaker identities of generated utterances, where lip sequences are randomly selected from testing speakers. Then the speaker representations of reference and generated utterances are extracted by using the speaker encoder used in the proposed VCVTS and visualized via t-SNE \cite{van2008visualizing} in the embedding space, as shown in Fig. \ref{spk-vis}. We observe that for both GRID and LRW, each reference utterance and its corresponding generated utterances can form a distinct cluster, this further shows that the proposed multi-speaker VTS system can use the speaker encoder to produce the speaker representation that effectively captures the speaker characteristics.

\vspace{-0.5em}
\section{Conclusions}
% \vspace{-0.5em}
We propose a novel multi-speaker VTS system, i.e., VCVTS, which can accurately reconstruct the spoken content from the lip motions and effectively control the speaker identity of generated speech. This is achieved by transferring the knowledge from VC, using the VQCPC-based content encoder of VC to guide the learning of a Lip2Ind network, and using the speaker representation produced by a speaker encoder to capture desired speaker characteristics. Compared with existing works, the proposed VTS system provides a more transparent mapping process from the lips to speech, quantitative and qualitative results show that state-of-the-art performance can be achieved based on various objective and subjective metrics under both constrained (e.g., GRID) and unconstrained (e.g., LRW) conditions. Our future work aims to study more challenging conditions for VTS including cross-domain and multi-lingual scenarios.

\section{ACKNOWLEDGEMENTS}
This research is supported partially by the HKSAR Research Grants Council's General Research Fund  (Ref Number 14208817) and also partially by the Centre for Perceptual and Interactive Intelligence, a CUHK InnoCentre.

\vfill\pagebreak

% \section{REFERENCES}
% \label{sec:refs}

% List and number all bibliographical references at the end of the
% paper. The references can be numbered in alphabetic order or in
% order of appearance in the document. When referring to them in
% the text, type the corresponding reference number in square
% brackets as shown at the end of this sentence \cite{C2}. An
% additional final page (the fifth page, in most cases) is
% allowed, but must contain only references to the prior
% literature.

% References should be produced using the bibtex program from suitable
% BiBTeX files (here: strings, refs, manuals). The IEEEbib.bst bibliography
% style file from IEEE produces unsorted bibliography list.
% -------------------------------------------------------------------------
\footnotesize
% \small
\begin{spacing}{0.9}
\bibliographystyle{IEEEbib}
\bibliography{strings,refs}
\end{spacing}

\end{document}